# A scheme for simulating multi-level phase change photonics materials

Yunzheng Wang[1,#*], Jing Ning[1,2#], Li Lu[1], Michel Bosman[2], Robert E. Simpson[1,*]

[1]Singapore University of Technology and Design (SUTD), 8 Somapah Road, 487372, Singapore

[2]Department of Materials Science and Engineering, National University of Singapore, 9 Engineering Drive 1, 117575, Singapore

Corresponding email: yunzheng_wang@sutd.edu.sg

Corresponding email: robert_simpson@sutd.edu.sg

[#] These authors contribute equally.

**Abstract**: Chalcogenide phase change materials (PCMs) have been extensively applied in data storage, and they are now being proposed for high resolution displays, holographic displays, reprogrammable photonics, and all-optical neural networks. These wide-ranging applications all exploit the radical property contrast between the PCMs' different structural phases, extremely fast switching speed, long-term stability, and low energy consumption. Designing PCM photonic devices requires an accurate model to predict the response of the device during phase transitions. Here, we describe an approach that accurately predicts the microstructure and optical response of phase change materials during laser induced heating. The framework couples the Gillespie Cellular Automata approach for modelling phase transitions with effective medium theory and Fresnel equations. The accuracy of the approach is verified by comparing the PCM's optical response and microstructure evolution with the results of nanosecond laser switching experiments. We anticipate that this approach to simulating the switching response of PCMs will become an important component for designing and simulating programmable photonics devices. The method is particularly important for predicting the multi-level optical response of PCMs, which is important for all-optical neural networks and PCM-programmable perceptrons.

# Introduction

Chalcogenide phase change materials (PCMs) exhibit extraordinarily large changes to their optical and electrical properties when switched between their local different bonding states[1-3]. The switching is reversible and can be cycled trillions of times[4]. This makes PCMs attractive not just for their existing application in electrical storage[5], but also for novel applications in photonics[6-9], micro-electromechanical systems[10], and tunable radio frequency devices[11].

Chalcogenide PCMs are particularly attractive for reprogrammable photonics because they exhibit a large optical refractive index change between amorphous and crystalline phases. For example, the real component of the refractive index of $Ge_2Sb_2Te_5$, which is the most famous PCMs, changes from ~4 to ~6 in the mid-infrared[12], making it useful for programming mid-infrared plasmonic metamaterials[13]. Chalcogenide PCMs tend to have a larger refractive index in both crystalline and amorphous phases than many other tunable photonics materials, and this makes them naturally suitable for designing all-dielectric metasurface-based devices[14]. Another appealing characteristic of chalcogenide PCMs is that the phase transition is extremely quick; indeed amorphisation takes just 5 ps[15] and crystallisation is possible in 700 ps[16]. Moreover, chalcogenide PCMs do not require a constant energy to hold their optical state once switched. This non-volatility makes them more energy efficient than the metal oxide PCMs, such as $VO_2$, which require a constant energy supply to hold their optical state. Importantly, the real part of the dielectric function of many telluride-based compositions is negative in the visible spectrum, which means that the crystalline material can behave like a plasmonic metal and could further enable new switchable plasmonic devices[17]. More recently, the multiple optical states of chalcogenide PCMs are being studied, developed, and applied in multi-level optical switches[18-20], a prerequisite for setting the weights in all-optical neural networks[21-23].

The biggest drawback of telluride-based PCMs is their high optical absorption at visible and near-infrared frequencies[12]. However, in the past few years, new PCMs, such as $Sb_2S_3$[24] and $Ge_2Sb_2Se_4Te_1$[25], have been developed to specifically program the response of visible and near-infrared photonics devices. Prototype optical switches[26,27], displays[28] and metasurfaces[29] have demonstrated their potential.

Designing programmable photonic devices requires an accurate model to describe the PCM switching behaviour. However, the optical properties of all materials depend on their respective crystal structures, and in the case of PCMs the crystal structure switches and transitions depending on the heating and quenching conditions. Even the crystallographic microstructure impacts optical scattering, and this may explain the large variation in measured optical constants of $Ge_2Sb_2Te_5$[12,30,31]. This variability issue is further confounded by most of the amorphous optical properties in literatures actually being for the as-deposited amorphous state, rather than the more technologically relevant melt-quenched amorphous state. To understand the evolution of a PCM photonics device optical response during switching, and the conditions that can lead to multi-level optical devices[18,19], it is important to know not only the initial and final states, but also how the optical constants evolve during nucleation, crystal growth, and melt-quenching. Therefore, a model that accurately predicts the optical properties during and after phase transition will be a key enabler of successful photonic device designs, and in particular phase change multi-level devices.

Approaches to modelling the optical change of PCMs during crystallisation and amorphisation range from simple but unrealistic descriptions of the refractive index switching at a specific temperature to *ab initio* descriptions of the electronic band structure dependence on the local atomic coordination[2] and computationally expensive density functional theory methods. However, semi-empirical methods that exploit molecular-level crystal nucleation, growth, and annihilation rates give a good compromise between accuracy, scaling, and performance. Recently, Meyer et al developed a multiphysics model that specifically describes the optical property change in a phase change metasurface perfect absorber using a phase field approach but no comparison with experiments was conducted to verify the multiphysics model[32].

Multiphysics models of $Ge_2Sb_2Te_5$ and other PCMs must accurately predict crystallisation behaviour across a time span ranging by 15 orders of magnitude because photonic memories must be stable at room temperature for years yet crystallise in nanoseconds at slightly elevated temperatures. This dichotomy is possible because the PCM viscosity depends non-linearly on temperature[33-36]. However, typical models that simulate the laser and electric heating pulses do not consider this non-linearity and thus they can only be used over narrow range of temperatures and heating rates.

The Gillespie Cellular Automata (GCA)-based phase-change model was developed by Ashwin to describe isothermal crystallisation of PCM alloys and phase transitions in electrical devices[37]. In the GCA model, each unit cell is considered as a PCM molecule with physically meaningful characteristics, and they can be directly related to microscopic density functional theory models. Therefore, the GCA approach was suggested as a bridge between hundred-atom-order density functional theory models and device-level performance models[37]. The GCA approach has been used to model electrical phase change memory devices and to study the PCM crystallisation process[38,39]. Although the electrical switching performance predicted from these models is in reasonable agreement with device measurements, the predicted crystallographic microstructure has not been verified by experiments. For photonics applications and especially multi-level photonics[19,20], it is important to know the influence of the crystal domain size and density on the PCM optical properties. Until now, however, a complete GCA-thermo-optical model that can be used to design photonics devices remains unreported.

The aim of this work is to develop an accurate framework for simulating PCM photonics devices. To quantify the accuracy, we must be able to simulate and measure the PCM microstructures and optical response of photonics devices. For this reason, we will apply the framework to a programmable thin-film optical stack. In what follows, we describe how combining nanosecond non-isothermal laser heating, non-linear and probabilistic GCA, effective medium theory and the Fresnel Equations can be used to accurately model the switching process of PCMs and the transient optical response of the programmable optical stack during nanosecond pulsed laser irradiation. Laser switching experiments and transmission electron microscopy (TEM) were performed to show that the model is highly accurate and can predict the final optical state, the crystallographic microstructure evolution, the transient optical response, and multilevel optical reflectivity and transmissivity which corresponds to partial crystallisation of the PCM film. Thus, this model is important because it can be used to optimise the PCM films, device structures, and laser pulse parameters to achieve multi-level switching. We foresee the model being widely applied to optimising the interconnection weights in all-optical neural network schemes[40], controllable metamaterial phase arrays[41,42], and displays[28,43].

For this reason, the code is publicly available at GitHub website[44], and we encourage others to use it.

## Results and Discussion

<u>Multiphysics Modelling</u>

The active structures in many PCM photonics devices consist of a PCM multilayer stack. Therefore, here we simulate the time-dependent optical response of such a stack during nanosecond laser heating. Note, however, the model can be easily adapted to other more complicated devices, such as waveguides and memories. The multi-physics scheme that we developed is shown in Figure 1. The optical response and microstructure of the stack used in experiments, can be easily measured, which makes this particular device ideal for validating the accuracy of the multi-physics model. The model starts by calculating the heat distribution due to laser heating. The temperature as a function of space and time was obtained by solving the heat conduction equation with an explicit finite difference method. Non-uniform meshing and time-step alternation strategies were adopted to increase the calculation efficiency. The phase-transition of the PCM molecules was simulated using a modified GCA model. In contrast to Ref[37], here, the probabilities of nucleation, growth and dissociation are directly deduced from the classical nucleation and growth theory, and these probabilities are combined with a piecewise viscosity model to simulate phase transitions. This step is important for handling the highly non-linear crystallisation rates which allow PCMs to crystallise from nanoseconds to years by only moderately changing the temperature. Due to the differences in dimensionality for the heat conduction and GCA grids, the 3D temperature distribution was re-sampled onto the 2D GCA grid, i.e., the calculation is assumed to occur in the top surface of the PCM film. This is acceptable because generally the thickness of PCM films in photonics devices is thinner than the PCM crystal grain size[45].

PCMs can exist in an intermediate phase, which simultaneously consists of crystalline and amorphous domains. Therefore, the crystallised fraction was computed as the ratio of crystalline molecules to the total number of molecules within the full width of half maximum (FWHM) of the Gaussian laser beam. The effective permittivity and optical constants of the crystalline grains within the amorphous matrix were calculated using the Lorentz-Lorenz relation and effective medium theory. The reflectivity and transmissivity of the device was updated by taking advantage of characteristic matrix method and Fresnel equations. The new absorption and reflection coefficients were further used to calculate the laser-induced heat distribution as the model iterates. Then, the multilevel responses of the PCM reprogrammable photonics device can be predicted. Further modelling details are described in the Methods section.

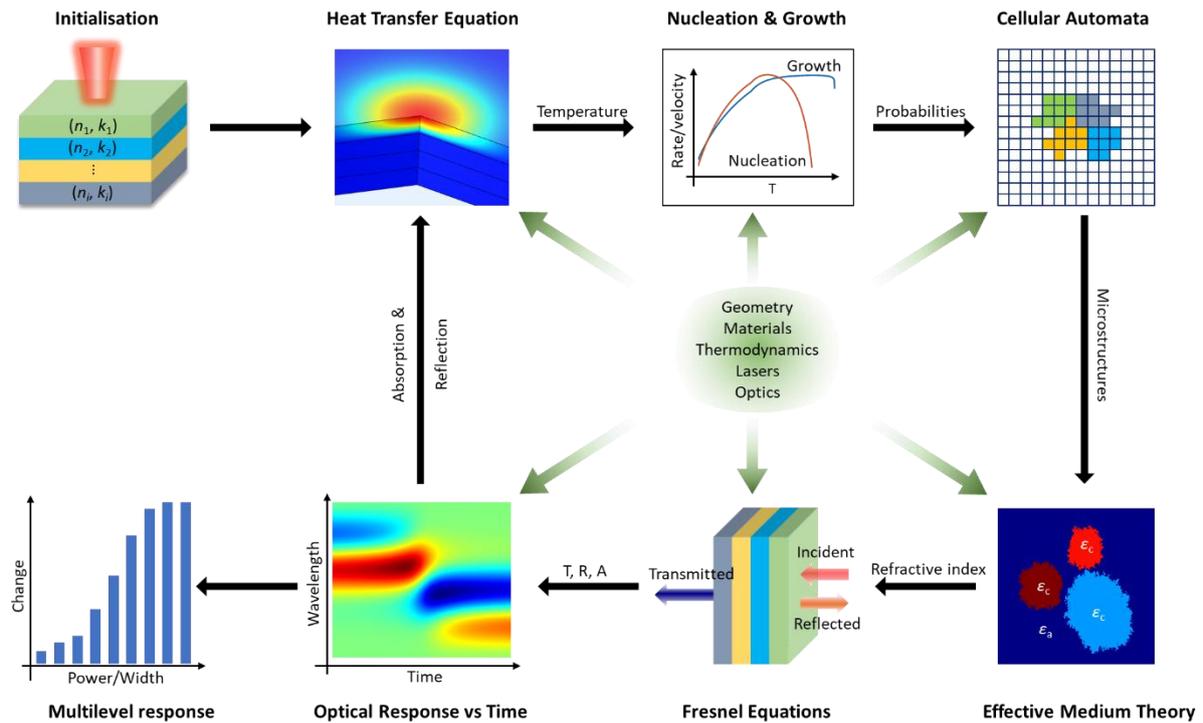

Figure 1. Schematic diagram of the proposed multi-physics GCA model.

Simulation

The multi-physics GCA model was applied to analyse the entire crystallisation process of a $Ge_2Sb_2Te_5$ sample under laser irradiation. The programmable thin film optical stack is shown in Figure 2(a). It consists of a 30 nm thick $Ge_2Sb_2Te_5$ film on top of a 50 nm thick silicon nitride ($Si_3N_4$) membrane. Again, this particular structure was simulated because of the ease at which the microstructure evolution can be measured. Indeed, the entire thickness of the stack is 80 nm, which is sufficiently thin for TEM imaging of the microstructure. Note that the current model can be easily adjusted to model crystallisation and amorphisation of any samples with a layered structure.

We simulated the crystallisation and melting processes of amorphous $Ge_2Sb_2Te_5$ with a focused Gaussian laser beam. The beam radius ($1/e^2$ intensity) was 600 nm and the peak power was changed from 0 mW to 5.0 mW, whilst the pulse duration was varied from 100 ns to 1000 ns. These parameters were chosen by considering the accessible parameters of our laser testing system[46], which will be used to compare the modelled and measured change in transmission due to the laser pulse. In this system, the temporal waveform of the laser pulse is an isosceles trapezoid function with rise and fall time of 8 ns. Due to the thinness of the $Ge_2Sb_2Te_5$ layer, the divergence of the laser beam in the $Ge_2Sb_2Te_5$ was ignored. At the 660 nm laser wavelength, the complex refractive indices of amorphous and crystalline $Ge_2Sb_2Te_5$ are $N_a = 4.36 - 1.79i$ and $N_c = 4.61 - 4.01i$, respectively[12].

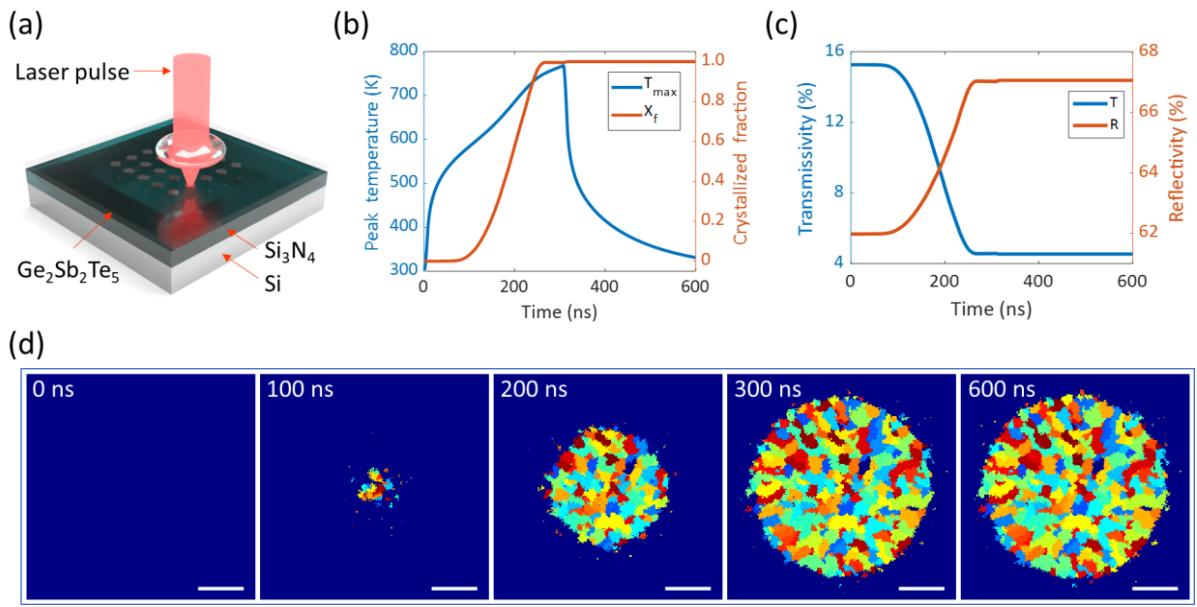

Figure 2. Schematic diagram of the $Ge_2Sb_2Te_5$ programmable optical stack (a) and simulated crystallisation at 3.0 mW (b-d). (b) The peak temperature ($T_{max}$) and crystallised fraction ($X_f$) versus time. (c) The transmissivity (T) and reflectivity (R) versus time. (d) The crystal microstructures of $Ge_2Sb_2Te_5$ film at different time. Scale bar is 200 nm.

The crystallisation process of the programmable optical device was simulated using a 3.0 mW and 300 ns laser pulse. Figure 2(b) and 2(c) show the peak temperature in the sample, the crystallised fraction, and the sample's transmissivity and reflectivity as a function of time, respectively. The peak temperature continuously rises during the laser pulse and then quickly quenches when the laser pulse ends. Figure 2(b) clearly shows that the crystallisation incubation time is 80 ns, at which time the peak temperature is 566 K. The crystallised fraction appears to stop growing after 260 ns because at this time the diameter of the crystallised area matches the FWHM of the laser spot. These characteristic times are also seen in the transmissivity and reflectivity curves, see Figure 2(c). Despite the crystallised fraction being close to 100% at 260 ns, the crystal is still growing. This effect is due to the crystallised fraction being computed under the FWHM of gaussian laser spot, which after 260 ns is smaller than crystal diameter. In addition, the rate that the peak temperature increases becomes slightly higher at 110 ns (see Figure S4 in Supporting Information (SI)), which is due to the higher single-pass absorption of the crystalised film in comparison to the amorphous background. The crystallographic microstructure of the $Ge_2Sb_2Te_5$ film at any time during crystallisation was also monitored. Figure 2(d) shows snapshots of the microstructure at 0, 100, 200, 300 and 600 ns. The blue background represents the amorphous phase, whereas other colours represent different orientations of crystalline grains. We see that the crystallised region is polycrystalline and consists of many small grains, indicating that the crystallisation of $Ge_2Sb_2Te_5$ is nucleation-dominated. Additionally, the crystallised region increases in size during the laser heat pulse and keeps almost constant once the laser pulse ends. This is important for understanding how to program a photonics device with a multi-level optical response. The diameter of the crystallised region was 810 ± 10 nm at 600 ns. Note that for these pulse conditions the grain size at the centre of the crystallised area is similar to that at the area's perimeter; we will see later that this is not always the case. The evolution of the microstructure is best seen using Movie 1 in SI.

Multilevel Switching Modelling

To show how our model can be used to find the conditions for multi-level switching, we used it to simulate the crystallisation dependence on laser power. Figure 3(a-c) shows the transient evolution of the crystallised fraction, reflectivity and transmissivity for 300 ns laser pulses with powers varying from 0 to 5.0 mW. The crystallised fraction, reflectivity and transmissivity exhibit a similar trend. We have also included the peak temperature evolution in Figure S5 of the SI. More importantly, we see that there are four distinct phases of evolution. Firstly, for laser powers below 2.0 mW, crystallisation cannot happen because the peak temperature is too low for the crystallites to grow. Indeed, less than 1% of the irradiated area has crystallised. Secondly, partial crystallisation is easily obtained for laser powers between 2.0 and 2.9 mW, which allows the crystallised percentage of material to be varied from 1% to 98%. This result is particularly significant for multi-level, or even programmable analogue optical states. Thirdly, complete crystallisation is realized within a power range of 3.0 to 3.8 mW; we see that 100% of the material that is irradiate by the laser has crystallised. Finally, when the laser power is greater than 3.8 mW, after the film crystallises, it melts and then quickly re-crystallises. This effect occurs because the dissociation probability of a GCA cell is higher than the growth probability for temperatures greater than 788 K (see Figure S2 in SI). Since crystallised percentage gradually decreases during dissociation (melting), the transmissivity increases and reflectivity decreases. In the model, we assumed that the molten phase of $Ge_2Sb_2Te_5$ has the same optical constants as the amorphous phase[47]. Recrystallisation of the molten phase takes up to twenty nanoseconds, and we see that the recrystallisation rate is higher than the crystallisation rate from an amorphous phase. This is due to the molten area being surrounded by a crystalline matrix; hence recrystallisation can occur from pre-existing crystal templates and there is no need to nucleate a crystal, which is time consuming. This is also reminiscent of high recrystallisation rates that are possible in $GeTe$-$Sb_2Te_3$ superlattice interfacial phase change materials, where crystallisation rates are high due to $Sb_2Te_3$ templating effects[48-50]. The melt-recrystallisation effect also verifies that pre-treating a PCM device can significantly increase the rate of phase transformations[51]. Note, the melt-quenched amorphous phase does not appear at high powers due to the low thermal conductivity of the floating $Si_3N_4$ membrane. Fig 2(d) showed the microstructure for crystallisation using 3.0 mW laser pulses, which didn't melt the $Ge_2Sb_2Te_5$. For comparison the crystal's microstructure evolution for 5 mW laser pulses is also given in Figure S6 in the SI. The Figure clearly shows melt-recrystallisation, and crystal growth from pre-existing templates causing large domain sizes at the centre of the irradiated mark.

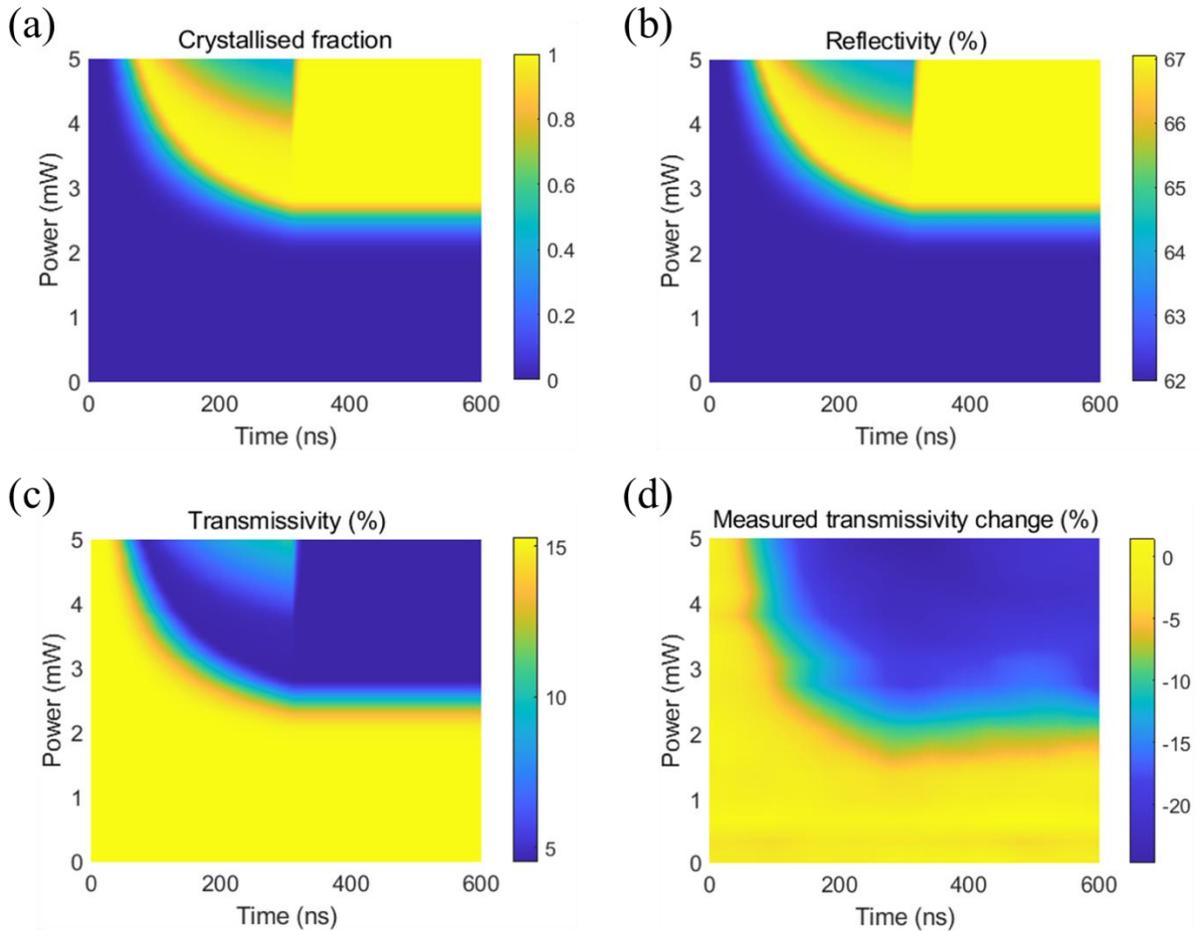

Figure 3. Transient evolution of (a) crystallised fraction, (b) reflectivity and (c) transmissivity at different laser power. (d) Evolution of measured relative transmissivity change during crystallisation.

So far, we have presented a model that predicts the time dependent change in optical transmissivity and reflectivity of a $Ge_2Sb_2Te_5$-based thin film programmable optical stack. We have shown that four switching regimes can take place which result in four distinct microstructural phases, and that the model can predict partial crystallisation which in turn can lead to multi-level optical switching. We now evaluate the accuracy of these predictions by laser switching a 30 nm thick $Ge_2Sb_2Te_5$ film on a $Si_3N_4$ membrane. Figure 3(d) shows the measured relative optical transmissivity change of the structure as a function of laser power and time for a fixed laser pulse width of 300 ns. The measured switching time and laser powers can be directly compared with the simulated results in Figure 3(c). There appears to be an excellent agreement. We see that the measured minimum laser power for crystallisation is $1.7 \pm 0.1$ mW, which is similar to the model. Partial crystallisation appears to be possible in a power range of 1.8 to 2.7 mW, and the model predicts 2.0 to 2.9 mW; once again there is a good agreement. Complete crystallisation is achieved for laser powers greater than 2.7 mW and the model predicts 3.0 mW. Qualitatively, we also see that the crystallisation time is shorter for higher power pulses. Indeed, at 5.0 mW, the crystallisation process starts at 40 ± 3 ns, which agrees well with the minimum predicted value (35 ± 3 ns). There is evidence of some small differences between the simulation and experiment results. The current model does not consider laser ablation. Ablation causes a substantial increase in the transmissivity, which is observed in the measurement when the laser power is raised above 5.5 mW (see Figure S9 & S10 in the SI). We also assumed that molten $Ge_2Sb_2Te_5$ had the same refractive index as amorphous $Ge_2Sb_2Te_5$. This means melting and recrystallisation can be distinguished in the transient transmissivity simulation. However, in the experiment, this effect wasn't observed. This difference might be due to molten $Ge_2Sb_2Te_5$ actually having similar optical properties to the crystalline $Ge_2Sb_2Te_5$.

Microstructure predictions

The modelled laser-crystallised marks quantitively and qualitatively agree with the measured laser crystallised marks. The crystal microstructures after crystallisation at different laser powers with the same 300 ns pulse width are shown in Figure 4(a). The crystallised regions can clearly be seen to increase in diameter from 0.68 um to 1.39 um as the laser power is increased from 2.8 mW to 4.9 mW. We see that for laser powers between 4.4 mW and 4.9 mW, the grain sizes are much larger at the centre of the crystallised area than at the perimeter, and this effect is caused by crystal growth from pre-existing crystalline surroundings without the need to nucleate. These larger grains were also verified by selected area electron diffraction (see Figure S11 in SI). By comparing simulations and measured TEM images, we know that for these laser powers, the centre of the irradiated area melts and then recrystallises. The larger grains at the centre of the mark do not appear at low power and short pulse widths because crystallisation occurs directly from the amorphous phase without melting. The diameters of the crystallised region at laser power larger than 3.3 mW are larger than the FWHM of laser beam (0.7 μm) because of the radial flow of heat outward from the laser spot. The predicted diameter of the crystallised marks shows excellent agreement with the measured diameters, see Figure 4(b).

Across the entire range of laser pulse powers studied, we see that the predicted diameter is within the measurement error. Therefore, the model accurately predicts the polycrystalline microstructure, grain size distribution, and the diameter of the crystallised region. This is useful because we can now be confident that this GCA model can predict the laser pulse parameters for multilevel switching.

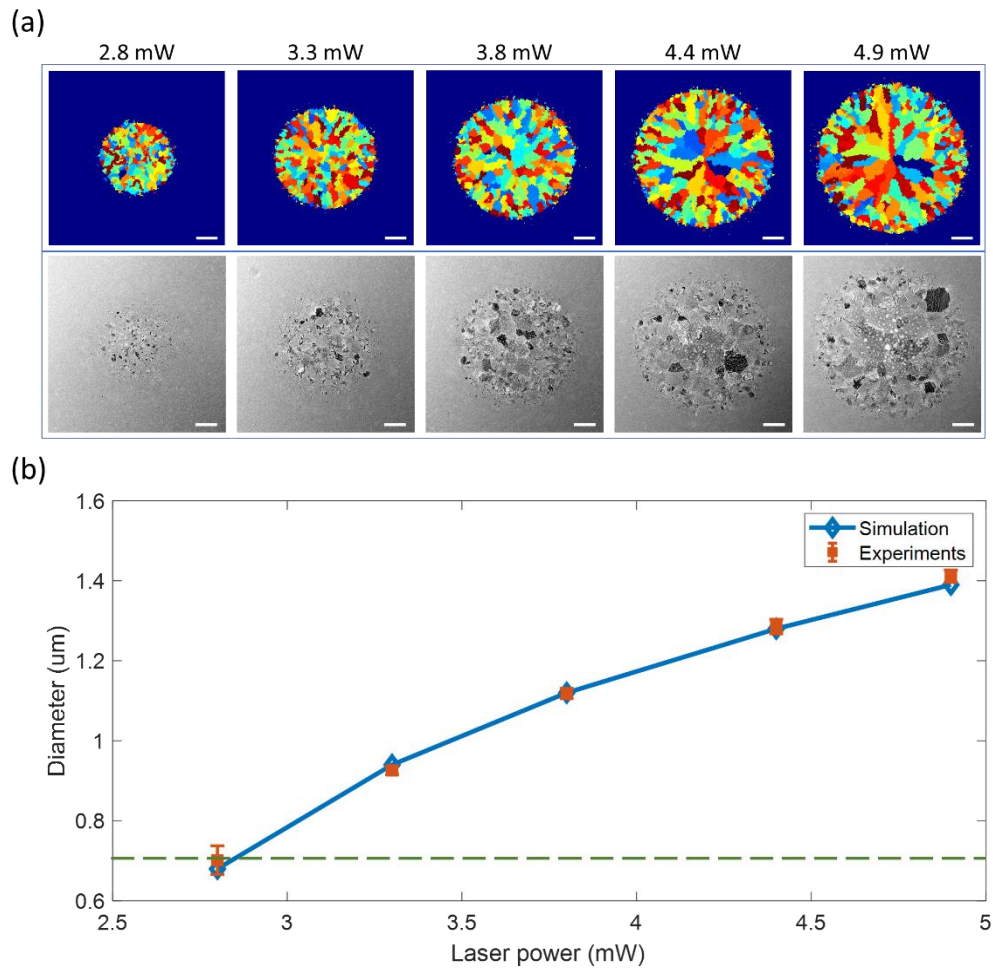

Figure 4. (a) Crystallographic microstructures at different laser power. First row: simulation; Second row: experiments. The scale bar is 200 nm. (b). Diameters of crystallised regions at different laser powers. The dashed green line indicates the FWHM of the laser beam.

Partial Crystallisation and multilevel switching

Multilevel switching can be achieved using a fixed pulse width and controlling the laser pulse power. Indeed, Rios et al used this method to control the output of a $Ge_2Sb_2Te_5$-programmable on-chip photonics device with four different transmission levels[20]. A second method to achieve multilevel switching involves using a fixed pulse power and changing the pulse duration (τ). We can see that this is possible by selecting a fixed power in Figure 3 and noting how the reflectivity and transmissivity change. Here, we simulated and measured partial crystallisation at a fixed laser power of 5.4 mW for different pulse widths. The total simulation time was 500 ns. Figure 5(a) shows the crystal percentage, reflectivity, and transmissivity after the applied laser pulse. Once again, there are four distinct steps to the phase transformation (also see Figure S7 in SI): (i) τ < 10 ns, crystallisation can be ignored due to too little heating; (ii) 10 ns< τ < 60 ns, partial crystallisation with the crystallised percentage being tuneable from 1% to 99%; (iii) 60 ns< τ < 100 ns, complete crystallisation is obtained; (iv) τ > 100 ns, melting and fast re-crystallisation appear during crystallisation. Importantly, we conclude that when the pulse duration is less than 60 ns, the reflectivity can be controlled using the pulse width.

The measured and simulated crystal microstructures after irradiating with 5.4 mW and varying pulse widths are shown in Figure 5(b). The diameter of the crystallised region increases with the laser pulse duration. Longer duration pulses caused melting and subsequent re-crystallisation into marks with large grains in the centre and smaller grains around the perimeter. Again, the agreement between the experiment and the simulation is good.

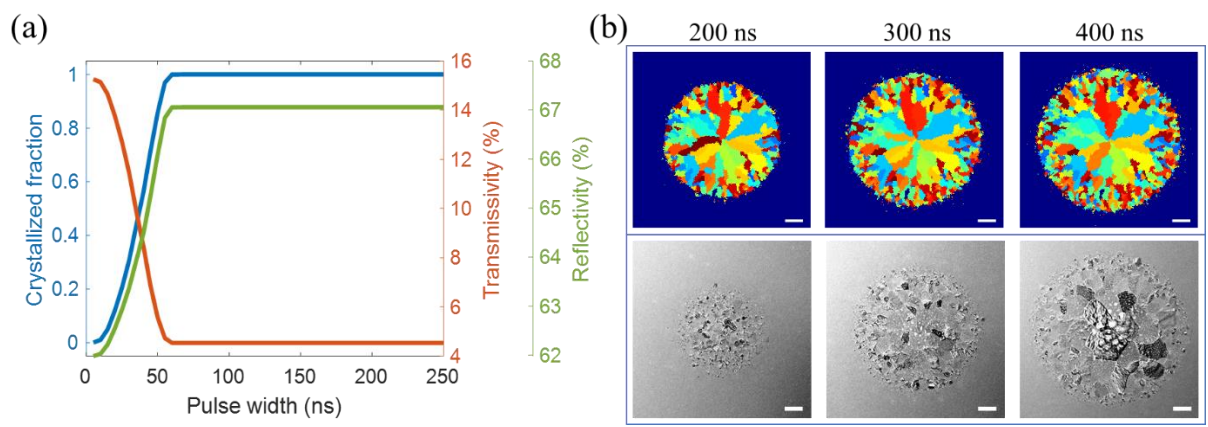

Figure 5. Mult-level switching by changing pulse width. (a) The crystal fraction, reflectivity, and transmissivity after 5.4 mW laser pulse with different widths. (b) Corresponding crystallographic microstructures. First row: simulation; Second row: experiments. The scale bar is 200 nm.

Multilevel switching laser parameter prediction

Achieving multiple optical states using $Ge_2Sb_2Te_5$ is challenging due to the rather abrupt optical response to heat pulses. Although multiple reflectivity levels have been achieved using picosecond laser pulses, generally reliably setting intermediate optical states with single nanosecond pulses is difficult. This is because the energy is delivered to the PCM over a similar period to that required for the phase transition, and because more heat is trapped in the surrounding structure. Hence, experimentally optimising the pulse conditions for different optical states is time consuming. Here, however, we found that the GCA Multi-physics modelling approach could accurately predict the pulse conditions for different optical reflectivity levels.

In the previous simulations and experiments, an implicit condition that both pump and probe lasers have the same beam size was used. Hence, a relatively narrow multilevel switching range was obtained due to the crystallised regions easily exceeding the FWHM of the probe laser. This issue can be mitigated by setting the probe laser spot size to be larger than the pump laser spot. We used the GCA multiphysics model to simulate the transmission change of the $Ge_2Sb_2Te_5$ thin-film programmable optical stack that is switched with different laser power and pulse widths. Here, the beam radii ($1/e^2$ intensity) of the pump and probe lasers were 0.8 μm and 1.1 μm, respectively. Figure 6(a) shows the normalised transmissivity change. Partial crystallisation and multi-level switching were realized for a power range of 2.4 mW to 5.0 mW, 2.6 times larger than results in Figure 3. Complete crystallisation was obtained for laser power > 5.0 mW. In experiments, the beam sizes were finely tuned to those used in the simulation. We can see from Figure 6(a) that the measured normalised transmission change has similar trend to that simulated. Similarly, the normalised transmission change induced by a 3.5 mW laser pulse with different widths was conducted in simulation and experiments, shown in Figure 6(b). The results also exemplify the accuracy of our multiphysics GCA model. The small differences may be caused by the deviation from the isosceles trapezoid of the temporal waveform of our laser pulses.

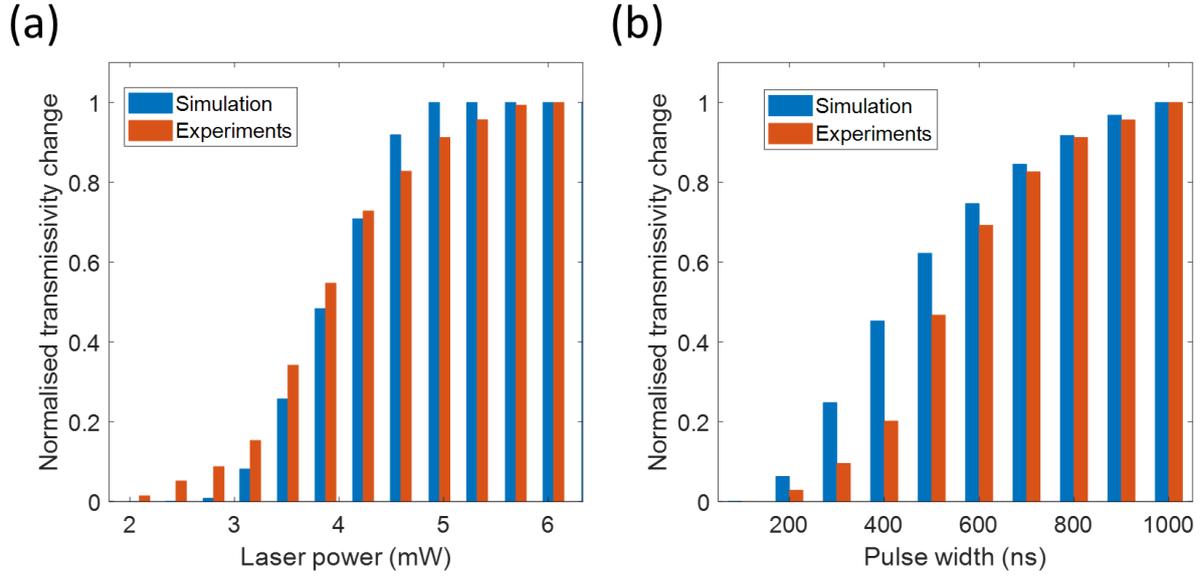

Figure 6. Multilevel switching performance. (a) Simulated and measured normalised transmissivity change at 300 ns laser pulses width different laser powers. (b) Simulated and measured normalised transmissivity change at 3.5 mW laser pulses width different widths.

Discussion

Crystallisation in PCMs is either nucleation or growth dominated. In nucleation driven materials, an incubation time is required for the material to create stable crystal nuclei. After incubation, a high density of nuclei form and each crystal domain tends to be small. Contrastingly, growth dominated materials do not readily nucleate, and tend to extend the crystalline regions from a nucleus that forms stochastically after a long incubation period. Consequently, the crystal domains in growth dominated materials tend to be larger than those of nucleation dominated materials. AgInSbTe and $Ge_2Sb_2Te_5$ are commonly referred to as growth and nucleation dominated materials, respectively[52]. However, it is also possible for $Ge_2Sb_2Te_5$ to exhibit a growth dominated crystallisation behaviour. At high temperature, the growth probability of $Ge_2Sb_2Te_5$ is much larger than the nucleation probability (see Figure S2 in SI), meaning that if pre-existing nucleation templates exist in the amorphous background, then $Ge_2Sb_2Te_5$ crystallisation becomes growth dominated. At high laser powers and long pulse widths, we see that the grain sizes at the centre of the heated region are clearly larger than those at the border. These larger grains are due to melting and subsequent re-crystallisation from the crystalline surroundings. This also indicates that the re-crystallisation process is growth dominated. From a device perspective, this finding is significant because if an amorphous mark is created within a crystalline surrounding, the optical constants of the material will not evolve by increasing the nuclei concentration but growing from pre-existing crystalline surroundings. This could be exploited to crystallise $Ge_2Sb_2Te_5$ in short times.

The modelling allowed us to successfully optimise the laser parameters to achieve more than 8 optical transmittivity levels in $Ge_2Sb_2Te_5$. Finding these parameters experimentally would be extremely time consuming due to the abrupt nature of the $Ge_2Sb_2Te_5$ phase transition. Other PCMs and multilayer PCM structures can make this response more gradual, and thus widen the window of pulses that can be used to achieve multilevel switching.[18,19] If one must use $Ge_2Sb_2Te_5$ in a multi-level switching photonics device, our simulations suggest that controlling both the pulse time and pulse power together will provide a wider window of possible partially crystallised states.

## Conclusion

A Gillespie cellular automata-based multiphysics model was developed to accurately predict the transient optical response of phase change material programmable photonics devices. We have used this model to simulate entire transient changes to the transmissivity and reflectivity of a programmable optical structure. By including the non-linear dependence of the viscosity with temperature, we were able to accurately predict the optical and microstructural properties of $Ge_2Sb_2Te_5$. Somewhat surprisingly, we found that laser pulses with relatively high intensities and long widths can cause $Ge_2Sb_2Te_5$ to exhibit growth dominated crystallisation, rather than nucleation dominated crystallisation, which is usual for $Ge_2Sb_2Te_5$. These results were confirmed by TEM measurements on laser switched $Ge_2Sb_2Te_5$. We also showed that the simulation tool can be used to accurately predict multilevel switching, and therefore we believe it will become a powerful tool to design and optimise new PCM programmable photonic devices. For this reason, we have made the code open for others to use[44].

Acknowledgements

This research was supported by the NLSM project (A18A7b0058) and support from A*STAR's microscopy facility is kindly acknowledged. Ms Ning is grateful for her Singapore Ministry of Education (MoE) PhD scholarship. The work was carried out under the auspices of the SUTD-MIT International Design Center (IDC).

# Methods

## Growth velocity and nucleation rate

Measurements of crystallisation kinetics for prototypical PCMs at low heating rates[53,54] showed that the growth velocity conforms to an Arrhenius exponential relation with reciprocal temperature, and typically this Arrhenius relationship is assumed in previously reported models of PCMs. Recently, however, ultrafast differential scanning colorimetry studies and laser and electrical heating experiments[33-36] showed that the growth velocity deviates strongly from the Arrhenius model at high temperatures due to the high fragility of the PCMs, which flattens the viscosity and growth velocity at high temperatures. The PCMs are in the super-cooled liquid state and the glass state when the temperature is above and below the glass transition temperature $T_{\text{glass}}$, respectively. The viscosity of this super-cooled liquid state is given by the Mauro-Yue-Ellison-Gupta-Allan model[55], whereas the viscosity of the glass state follows the Arrhenius relationship (see S2 in SI).

The crystal growth velocity (i.e., the growth rate of the radius of a crystal cluster per unit time) is modelled as

$$v_g = \frac{4rk_BT}{3\pi\lambda_j^2 R_h \eta(T)} \left[1 - \exp\left(-\frac{\Delta g}{k_BT}\right)\right] \quad (1)$$

where, $r$ and $R_h$ are the atomic radius and the hydrodynamic radius, respectively, $k_B$ is the Boltzmann constant, $T$ is the temperature, $\lambda_j$ is the diffusional jump distance, $\eta(T)$ is the viscosity, $\Delta g$ is the bulk Gibbs energy difference per monomer between the liquid and crystalline phase. According to Thompson-Spaepen model[56], $\Delta g$ is calculated as

$$\frac{\Delta g}{v_m} = \Delta H_f \frac{T_m - T}{T_m} \frac{2T}{T_m + T} \tag{2}$$

where, $v_m$ is the volume of one monomer, $\Delta H_f$ is the enthalpy of fusion at the melting temperature $T_m$. According to the classical growth theory, the two terms of Eq. (1) correspond to the growth ($v_{gr}$) and dissociation ($v_{di}$) velocity, respectively, which means that Eq. (1) represents the net growth velocity.

Based on the classical nucleation theory, the nucleation rate is written as[57]

$$I^{ss} = (4/v_m)\gamma n_c^{2/3} \sqrt{\frac{\Delta g}{6\pi k_B T n_c}} \exp\left(-\frac{\Delta G_c}{k_B T}\right) \tag{3}$$

where, $n_c = f(\theta) \cdot 32\pi/3 \cdot v_m^2 \sigma^3/\Delta g^3$ is the number of monomers contained in the critical crystal clusters, $\Delta G_c = f(\theta) \cdot 16\pi/3 \cdot v_m^2 \sigma^3/\Delta g^2$ is the critical energy barrier for nucleation, $\sigma$ is the interfacial energy, $f(\theta) = (2-3\cos\theta + \cos^3\theta)/4$ is the factor that indicates the degree of the heterogeneous nucleation, $\theta$ is the wetting angle, $\gamma$ is the molecular jump frequency which is modelled by the Stokes-Einst31ein relation[57] as $\gamma = k_B T/3\pi\lambda_j^3\eta$. Here, the nucleation rate means the number of newly generated nuclei per unit volume and time. The typical parameters for $Ge_2Sb_2Te_5$ are given in Table S1 in SI.

**Modified GCA Model**

In the previously reported GCA models[37], the PCM is divided into a two-dimensional matrix. Each site $(i, j)$ has a phase variable $r_{ij}$ to indicate whether it is in crystalline ($r_{ij} = 1$) or amorphous ($r_{ij} = 0$) phase, and an orientation variable $\phi_{ij}$ with a range of $[0, \pi)$ to represent the infinite possible orientations of the crystalline grain. Three possible events can impact the state of each site: (a) *Nucleation*; an amorphous cell transforms into a crystalline cell to form a new crystal nucleus with a probability of $P_{nu}(i, j)$. Then, $r_{ij} = 1$ and $\phi_{ij}$ is the orientation of the nucleus; (b) *Growth*; an amorphous cell grows onto an adjacent crystallite with an orientation $\psi$ with a probability of $P_{gr}(i, j, \psi)$. Then, $r_{ij} = 1$ and $\phi_{ij} = \psi$; (c) *Dissociation*; a crystalline site detaches from its parent crystallite and becomes amorphous phase with a probability of $P_{di}(i, j)$. Then, $r_{ij} = 0$ and $\phi_{ij}$ becomes a random value meaning a random orientation. The GCA model uses a stochastic Gillespie-type algorithm to simultaneously compute the time step and the state of the site using the probabilities of all events.

In contrast to Ref[37], here the probabilities of the three events were directly deduced from the above growth velocity and nucleation rate equations, and this allowed accurate predictions of the phase transition across a wide range of temperatures. Considering the influence of neighbours on a site as a linear approximation, the nucleation probability $P_{nu}(i, j)$ for an amorphous site to nucleate is written as

$$P_{nu}(i,j) = n_{ij}^{am} I^{ss} n_c v_m = \begin{cases} 4n_{ij}^{am} \gamma n_c^{5/3} \sqrt{\dfrac{\Delta g}{6\pi k_B T n_c}} \exp\left(-\dfrac{\Delta G_c}{k_B T}\right), & \text{if } r_{ij} = 0 \\ 0, & \text{if } r_{ij} = 1 \end{cases} \quad (4)$$

where, $n_{ij}^{am}$ is the number of amorphous neighbors of a site. The growth probability $P_{gr}(i, j, \psi)$ for an amorphous site to crystallise according to an adjacent crystallite with an orientation $\psi$ is written as

$$P_{gr}(i,j,\psi) = n_{ij}^{\psi} \dfrac{v_{gr}}{D} = \begin{cases} \dfrac{4 n_{ij}^{\psi} r k_B T}{3\pi \lambda_j^2 D R_h \eta(T)}, & \text{if } r_{ij} = 0 \\ 0, & \text{if } r_{ij} = 1 \end{cases} \quad (5)$$

and the dissociation probability $P_{di}(i, j)$ for a crystalline site to become amorphous is written as

$$P_{di}(i,j) = (n_{ij} - n_{ij}^{\psi}) \dfrac{v_{di}}{D} = \begin{cases} 0, & \text{if } r_{ij} = 1 \\ \dfrac{4(n_{ij} - n_{ij}^{\psi}) r k_B T}{3\pi \lambda_j^2 D R_h \eta(T)} \exp\left(-\dfrac{\Delta g}{k_B T}\right), & \text{if } r_{ij} = 0 \end{cases} \quad (6)$$

where, $n_{ij}$ is the total number of neighbours of a site which is equal to 8 in the two-dimensional matrix, $n_{ij}^{\psi}$ is the number of crystalline neighbors with an orientation $\psi$ of a site, and $D$ is the distance between two adjacent sites.

## laser-induced heat distribution

A laser pulse can increase the temperature of PCMs and induce crystallisation, melting, and/or amorphisation. The transient temperature profile is obtained by calculating the unsteady heat conduction equation,

$$\rho c \frac{\partial T(x,y,z,t)}{\partial t} = k\nabla^2 T(x,y,z,t) + Q(x,y,z,t) \tag{7}$$

where, $T(x,y,z,t)$ is the temperature at a location of $(x,y,z)$ and a certain time $t$, $\rho$ is the density, $c$ is the specific heat capacity, $k$ is the heat conduction coefficient, $Q(x,y,z,t)$ is the heat source brought by the laser pulse. Considering the laser beam's Gaussian transverse profile, the laser-induced heat is modelled as

$$Q(x,y,z,t) = \frac{2P_{in}\alpha}{\pi\omega^2}(1-R)e^{-2\frac{x^2+y^2}{\omega^2}}e^{-\alpha z}f(t) \tag{8}$$

where, $P_{in}$ is the optical power, $\alpha$ is the absorption coefficient, $\omega$ is the gaussian beam radius, $R$ is the reflectivity, and $f(t)$ is the temporal waveform. An isosceles trapezoid function with rising and falling edges of 8 ns was assumed for $f(t)$ in simulations to reflect the laser waveforms in experiments.

An explicit finite difference method was used to compute Eq. (7) to obtain the transient temperature profile of a given sample structure. However, the maximum time step is limited to $\Delta t_{max} = \rho c h^2/6k$ in three-dimensions in order to satisfy the stability requirements, where $h$ is the space step. To increase the efficiency of the calculation while maintaining high precision, a non-uniform meshing technique was exploited by dividing more grid points at laser heating regions and fewer grid points at regions far from the laser beam (see S4 in SI). Additionally, a time-step alternation strategy was adopted to further accelerate the calculation (see S5 in SI).

## Optical Model

During laser pulse irradiation on the PCM sample, the temperature rapidly increases, and many PCM sites transform to new phases, leading to the formation of an intermediate phase which simultaneously consists of crystalline and amorphous phases. According to the effective medium theory, the effective permittivity $\varepsilon_{eff}(\lambda)$ of this intermediate PCM is calculated based on Lorentz-Lorenz relation[58],

$$\frac{\varepsilon_{eff}(\lambda)-1}{\varepsilon_{eff}(\lambda)+2} = X \times \frac{\varepsilon_c(\lambda)-1}{\varepsilon_c(\lambda)+2} + (1-X) \times \frac{\varepsilon_a(\lambda)-1}{\varepsilon_a(\lambda)+2} \qquad (9)$$

where, $X$ is the crystallised fraction which is defined as the ratio of crystalline sites to the total sites. $\varepsilon_c(\lambda)$ and $\varepsilon_a(\lambda)$ are the wavelength-dependent dielectric functions for the crystalline and amorphous phases, respectively. Then, the effective refractive index ($n_{eff}$) and extinction coefficient ($k_{eff}$) are given by

$$n_{eff} = \sqrt{\frac{\sqrt{(\varepsilon_1+\varepsilon_2)^2}+\varepsilon_1}{2}}, \quad k_{eff} = \sqrt{\frac{\sqrt{(\varepsilon_1+\varepsilon_2)^2}-\varepsilon_1}{2}} \qquad (10)$$

where, $\varepsilon_1$ and $\varepsilon_2$ are the real and imaginary parts of $\varepsilon_{eff}(\lambda)$, respectively.

Using the characteristic matrix approach and the Fresnel equations, the reflectivity and transmissivity of a PCM sample at normal incidence are calculated as[59]

$$R = \left|\frac{m_{11}+n_s m_{12}-m_{21}-n_s m_{22}}{m_{11}+n_s m_{12}+m_{21}+n_s m_{22}}\right|^2, \quad T = \frac{4\operatorname{Re}(n_s)}{|m_{11}+n_s m_{12}+m_{21}+n_s m_{22}|^2} \qquad (11)$$

where, $n_s$ is the complex refractive index of substrate, Re($n_s$) means the real part of $n_s$, $m_{ij}$ ($i,j$ = 1, 2) are the elements of the characteristic matrix of the PCM sample.

## Sample preparation and characterization

A 50 nm thick silicon nitride membrane window (TED Pella, No. 21509) supported by a 0.5 mm silicon substate was chosen as substrate due to the window's good electron transmission characteristics. A 30 nm thick amorphous $Ge_2Sb_2Te_5$ film was deposited on the substrate by radio frequency magnetron sputtering from a $Ge_2Sb_2Te_5$ target at 30 W for 67 seconds in an argon atmosphere at a pressure of 0.5 Pa. Our inhouse-built static tester[46] was used to make crystallisation marks on the $Ge_2Sb_2Te_5$ film. In short, the system consists of a low-power 638 nm probe laser and a relatively high-power 660 nm pump laser. The system can simultaneously measure the reflection and transmission of the probe laser from the sample whilst pump laser pulses heat the sample. The focused laser spot has a beam size of 0.6 μm (1/e² intensity) on the sample. Here, we used the static tester to make a matrix of crystallisation marks under different pulse widths and intensities. Simultaneously, the transient transmission change was measured. The absolute value of the relative transmissivity change cannot be directly compared with the simulated transmissivity change because there is a DC offset on the amplified detector and because the photodetector does not collect all the transmitted light.

The size and microstructure of the crystallised marks were then measured by Transmission Electron Microscopy (FEI Titan with Gatan OneView camera) using an acceleration voltage of 200 kV and a rather small 40 μm objective aperture to enhance the contrast of the crystallised regions. The beam current was kept sufficiently low so that the PCM state was unaffected by the heating. A schematic of the TEM sample and its preparation is given in Figure S8 in the SI.

# References

1. Lee, T. H. & Elliott, S. R. Chemical Bonding in Chalcogenides: The Concept of Multicenter Hyperbonding. *Advanced Materials* **2020,** *32* (28), 2000340.
2. Martinez, J. C., Lu, L., Ning, J., Dong, W., Cao, T. & Simpson, R. E. The Origin of Optical Contrast in Sb2Te3-Based Phase-Change Materials. *physica status solidi (b)* **2020,** *257* (1), 1900289.
3. Zhu, M., Cojocaru-Miredin, O., Mio, A. M., Keutgen, J., Kupers, M., Yu, Y., Cho, J. Y., Dronskowski, R. & Wuttig, M. Unique Bond Breaking in Crystalline Phase Change Materials and the Quest for Metavalent Bonding. *Advanced Materials* **2018,** *30* (18).
4. Kim, W., BrightSky, M., Masuda, T., Sosa, N., Kim, S., Bruce, R., Carta, F., Fraczak, G., Cheng, H. Y., Ray, A.et al. 2016 IEEE International Electron Devices Meeting (IEDM), 2016; p 4.2.1.
5. Hoddeson, L. & Garrett, P. The discovery of Ovshinsky switching and phase-change memory. *Physics Today* **2018,** *71* (6), 44.
6. Cao, T., Wang, R., Simpson, R. E. & Li, G. Photonic Ge-Sb-Te phase change metamaterials and their applications. *Progress in Quantum Electronics* **2020,** *74*, 100299.
7. Rudé, M., Pello, J., Simpson, R. E., Osmond, J., Roelkens, G., Tol, J. J. G. M. v. d. & Pruneri, V. Optical switching at 1.55 μm in silicon racetrack resonators using phase change materials. *Appl Phys Lett* **2013,** *103* (14), 141119.
8. Rudé, M., Simpson, R. E., Quidant, R., Pruneri, V. & Renger, J. Active Control of Surface Plasmon Waveguides with a Phase Change Material. *ACS Photonics* **2015,** *2* (6), 669.
9. Wang, Q., Rogers, E. T. F., Gholipour, B., Wang, C.-M., Yuan, G., Teng, J. & Zheludev, N. I. Optically reconfigurable metasurfaces and photonic devices based on phase change materials. *Nature Photonics* **2016,** *10* (1), 60.

10. Wilhelm, E., Richter, C. & Rapp, B. E. Phase change materials in microactuators: Basics, applications and perspectives. *Sensors and Actuators A: Physical* **2018,** *271*, 303.
11. Champlain, J. G., Ruppalt, L. B., Guyette, A. C., El-Hinnawy, N., Borodulin, P., Jones, E., Young, R. M. & Nichols, D. Examination of the temperature dependent electronic behavior of GeTe for switching applications. *J Appl Phys* **2016,** *119* (24), 244501.
12. Chew, L. T., Dong, W., Liu, L., Zhou, X., Behera, J., Liu, H., Sreekanth, K., Mao, L., Cao, T., Yang, J.et al. *Chalcogenide active photonics*; SPIE, 2017.
13. Dong, W., Qiu, Y., Zhou, X., Banas, A., Banas, K., Breese, M. B. H., Cao, T. & Simpson, R. E. Tunable Mid-Infrared Phase-Change Metasurface. *Adv Opt Mater* **2018,** *6* (14), 1701346.
14. Leitis, A., Heßler, A., Wahl, S., Wuttig, M., Taubner, T., Tittl, A. & Altug, H. All-Dielectric Programmable Huygens' Metasurfaces. *Advanced Functional Materials* **2020,** *30* (19), 1910259.
15. Waldecker, L., Miller, T. A., Rudé, M., Bertoni, R., Osmond, J., Pruneri, V., Simpson, R. E., Ernstorfer, R. & Wall, S. Time-domain separation of optical properties from structural transitions in resonantly bonded materials. *Nature Materials* **2015,** *14* (10), 991.
16. Rao, F., Ding, K., Zhou, Y., Zheng, Y., Xia, M., Lv, S., Song, Z., Feng, S., Ronneberger, I., Mazzarello, R.et al. Reducing the stochasticity of crystal nucleation to enable subnanosecond memory writing. *Science* **2017,** *358* (6369), 1423.
17. Gholipour, B., Karvounis, A., Yin, J., Soci, C., MacDonald, K. F. & Zheludev, N. I. Phase-change-driven dielectric-plasmonic transitions in chalcogenide metasurfaces. *NPG Asia Materials* **2018,** *10* (6), 533.
18. Meng, Y., Behera, J. K., Ke, Y., Chew, L., Wang, Y., Long, Y. & Simpson, R. E. Design of a 4-level active photonics phase change switch using VO2 and Ge2Sb2Te5. *Appl Phys Lett* **2018,** *113* (7), 071901.
19. Meng, Y., Behera, J. K., Wen, S., Simpson, R. E., Shi, J., Wu, L., Song, Z., Wei, J. & Wang, Y. Ultrafast Multilevel Optical Tuning with CSb2Te3 Thin Films. *Adv Opt Mater* **2018,** *6* (17), 1800360.
20. Ríos, C., Stegmaier, M., Hosseini, P., Wang, D., Scherer, T., Wright, C. D., Bhaskaran, H. & Pernice, W. H. P. Integrated all-photonic non-volatile multi-level memory. *Nature Photonics* **2015,** *9* (11), 725.
21. Li, X., Youngblood, N., Ríos, C., Cheng, Z., Wright, C. D., Pernice, W. H. P. & Bhaskaran, H. Fast and reliable storage using a 5 bit, nonvolatile photonic memory cell. *Optica* **2019,** *6* (1), 1.
22. Wuttig, M., Bhaskaran, H. & Taubner, T. Phase-change materials for non-volatile photonic applications. *Nature Photonics* **2017,** *11* (8), 465.
23. Teo, T. Y., Lu, L. & Simpson, R. E. Reconfigurable multi-bit phase change material silicon photonics directional couplers. *arXiv preprint arXiv:2106.01169* **2021**.
24. Dong, W., Liu, H., Behera, J. K., Lu, L., Ng, R. J. H., Sreekanth, K. V., Zhou, X., Yang, J. K. W. & Simpson, R. E. Wide Bandgap Phase Change Material Tuned Visible Photonics. *Advanced Functional Materials* **2019,** *29* (6), 1806181.

25. Zhang, Y., Li, J., Chou, J., Fang, Z., Yadav, A., Lin, H., Du, Q., Michon, J., Han, Z., Huang, Y. et al. Conference on Lasers and Electro-Optics, San Jose, California, 2017; p JTh5C.4.
26. Zhang, Y., Chou, J. B., Li, J., Li, H., Du, Q., Yadav, A., Zhou, S., Shalaginov, M. Y., Fang, Z., Zhong, H. et al. Broadband transparent optical phase change materials for high-performance nonvolatile photonics. *Nature Communications* **2019**, *10* (1), 4279.
27. Delaney, M., Zeimpekis, I., Lawson, D., Hewak, D. W. & Muskens, O. L. A New Family of Ultralow Loss Reversible Phase-Change Materials for Photonic Integrated Circuits: Sb2S3 and Sb2Se3. *Advanced Functional Materials* **2020**, *30* (36), 2002447.
28. Liu, H., Dong, W., Wang, H., Lu, L., Ruan, Q., Tan, Y. S., Simpson, R. E. & Yang, J. K. W. Rewritable color nanoprints in antimony trisulfide films. *Science Advances* **2020**, *6* (51), eabb7171.
29. Zhang, Y., Fowler, C., Liang, J., Azhar, B., Shalaginov, M. Y., Deckoff-Jones, S., An, S., Chou, J. B., Roberts, C. M., Liberman, V. et al. Electrically reconfigurable non-volatile metasurface using low-loss optical phase-change material. *Nature Nanotechnology* **2021**, DOI:10.1038/s41565-021-00881-9 10.1038/s41565-021-00881-9.
30. Shportko, K., Kremers, S., Woda, M., Lencer, D., Robertson, J. & Wuttig, M. Resonant bonding in crystalline phase-change materials. *Nature Materials* **2008**, *7* (8), 653.
31. Gemo, E., Kesava, S. V., Ruiz De Galarreta, C., Trimby, L., García-Cuevas Carrillo, S., Riede, M., Baldycheva, A., Alexeev, A. & Wright, C. D. Simple technique for determining the refractive index of phase-change materials using near-infrared reflectometry. *Opt. Mater. Express* **2020**, *10* (7), 1675.
32. Meyer, S., Tan, Z. Y. & Chigrin, D. N. Multiphysics simulations of adaptive metasurfaces at the meta-atom length scale. *Nanophotonics* **2020**, *9* (3), 675.
33. Orava, J., Greer, A. L., Gholipour, B., Hewak, D. W. & Smith, C. E. Characterization of supercooled liquid Ge_2Sb_2Te_5 and its crystallization by ultrafast-heating calorimetry. *Nat Mater* **2012**, *11* (4), 279.
34. Salinga, M., Carria, E., Kaldenbach, A., Bornafft, M., Benke, J., Mayer, J. & Wuttig, M. Measurement of crystal growth velocity in a melt-quenched phase-change material. *Nat Commun* **2013**, *4*.
35. Sebastian, A., Le Gallo, M. & Krebs, D. Crystal growth within a phase change memory cell. *Nature Communications* **2014**, *5* (1), 4314.
36. Chen, B., ten Brink, G. H., Palasantzas, G. & Kooi, B. J. Crystallization Kinetics of GeSbTe Phase-Change Nanoparticles Resolved by Ultrafast Calorimetry. *The Journal of Physical Chemistry C* **2017**, *121* (15), 8569.
37. Ashwin, P., Patnaik, B. S. V. & Wright, C. D. Fast simulation of phase-change processes in chalcogenide alloys using a Gillespie-type cellular automata approach. *J Appl Phys* **2008**, *104* (8), 084901.
38. Wright, C. D., Hosseini, P. & Diosdado, J. A. V. Beyond von-Neumann Computing with Nanoscale Phase-Change Memory Devices. *Advanced Functional Materials* **2013**, *23* (18), 2248.
39. Diosdado, J. A. V., Ashwin, P., Kohary, K. I. & Wright, C. D. Threshold switching via electric field induced crystallization in phase-change memory devices. *Appl Phys Lett* **2012**, *100* (25), 253105.

40. Miscuglio, M., Adam, G. C., Kuzum, D. & Sorger, V. J. Roadmap on material-function mapping for photonic-electronic hybrid neural networks. *Apl Mater* **2019,** *7* (10), 100903.
41. de Galarreta, C. R., Alexeev, A. M., Au, Y.-Y., Lopez-Garcia, M., Klemm, M., Cryan, M., Bertolotti, J. & Wright, C. D. Nonvolatile Reconfigurable Phase-Change Metadevices for Beam Steering in the Near Infrared. *Advanced Functional Materials* **2018,** *28* (10), 1704993.
42. de Galarreta, C. R., Sinev, I., Alexeev, A. M., Trofimov, P., Ladutenko, K., Carrillo, S. G.-C., Gemo, E., Baldycheva, A., Nagareddy, V. K. & Bertolotti, J. All-dielectric silicon/phase-change optical metasurfaces with independent and reconfigurable control of resonant modes. *arXiv preprint arXiv:1901.04955* **2019**.
43. Hosseini, P., Wright, C. D. & Bhaskaran, H. An optoelectronic framework enabled by low-dimensional phase-change films. *Nature* **2014,** *511* (7508), 206.
44. Wang, Y., Ning, J., Lu, L., Bosman, M. & Simpson, R. E. Multiphysics GCA codes. (2021), https://github.com/YunzhengWang/multiphysics_GCA.
45. Ohshima, N. Crystallization of germanium–antimony–tellurium amorphous thin film sandwiched between various dielectric protective films. *J Appl Phys* **1996,** *79* (11), 8357.
46. Behera, J. K., Zhou, X., Tominaga, J. & Simpson, R. E. Laser switching and characterisation of chalcogenides: systems, measurements, and applicability to photonics [Invited]. *Opt. Mater. Express* **2017,** *7* (10), 3741.
47. Kuwahara, M., Endo, R., Fukaya, T., Shima, T., Iwanabe, Y., Fons, P., Tominaga, J. & Susa, M. A Reversible Change of Reflected Light Intensity between Molten and Solidified Ge–Sb–Te Alloy. *Japanese Journal of Applied Physics* **2007,** *46* (No. 36), L868.
48. Zhou, X., Behera, J. K., Lv, S., Wu, L., Song, Z. & Simpson, R. E. Avalanche atomic switching in strain engineered Sb2Te3–GeTe interfacial phase-change memory cells. *Nano Futures* **2017,** *1* (2), 025003.
49. Simpson, R. E., Fons, P., Kolobov, A. V., Krbal, M. & Tominaga, J. Enhanced crystallization of GeTe from an Sb2Te3 template. *Appl Phys Lett* **2012,** *100* (2), 021911.
50. Simpson, R. E., Fons, P., Kolobov, A. V., Fukaya, T., Krbal, M., Yagi, T. & Tominaga, J. Interfacial phase-change memory. *Nature Nanotechnology* **2011,** *6* (8), 501.
51. Lee, B.-S., Burr, G. W., Shelby, R. M., Raoux, S., Rettner, C. T., Bogle, S. N., Darmawikarta, K., Bishop, S. G. & Abelson, J. R. Observation of the Role of Subcritical Nuclei in Crystallization of a Glassy Solid. *Science* **2009,** *326* (5955), 980.
52. Zhang, W., Mazzarello, R., Wuttig, M. & Ma, E. Designing crystallization in phase-change materials for universal memory and neuro-inspired computing. *Nat Rev Mater* **2019,** *4* (3), 150.
53. Kalb, J., Spaepen, F. & Wuttig, M. Atomic force microscopy measurements of crystal nucleation and growth rates in thin films of amorphous Te alloys. *Appl Phys Lett* **2004,** *84* (25), 5240.
54. Kooi, B. J., Pandian, R., De Hosson, J. T. M. & Pauza, A. In situ transmission electron microscopy study of the crystallization of fast-growth doped SbxTe alloy films. *Journal of Materials Research* **2005,** *20* (7), 1825.

55. Mauro, J. C., Yue, Y., Ellison, A. J., Gupta, P. K. & Allan, D. C. Viscosity of glass-forming liquids. *Proceedings of the National Academy of Sciences* **2009,** *106* (47), 19780.
56. Thompson, C. V. & Spaepen, F. On the approximation of the free energy change on crystallization. *Acta Metallurgica* **1979,** *27* (12), 1855.
57. Senkader, S. & Wright, C. D. Models for phase-change of Ge2Sb2Te5 in optical and electrical memory devices. *J Appl Phys* **2004,** *95* (2), 504.
58. Qu, Y., Li, Q., Du, K., Cai, L., Lu, J. & Qiu, M. Dynamic Thermal Emission Control Based on Ultrathin Plasmonic Metamaterials Including Phase-Changing Material GST. *Laser & Photonics Reviews* **2017,** *11* (5), 1700091.
59. Larouche, S. & Martinu, L. OpenFilters: open-source software for the design, optimization, and synthesis of optical filters. *Appl. Opt.* **2008,** *47* (13), C219.